\documentclass[review]{elsarticle}
\makeatletter
\def\ps@pprintTitle{%
 \let\@oddhead\@empty
 \let\@evenhead\@empty
 \def\@oddfoot{}%
 \let\@evenfoot\@oddfoot}
\makeatother
\usepackage{hyperref,graphicx}
\usepackage[normalem]{ulem}
\usepackage[margin=1.0in]{geometry} 
\usepackage{multirow}
\usepackage{subfigure}
\usepackage{amsfonts}
\usepackage{amssymb}
\usepackage{amsmath,bm}
\usepackage{booktabs, threeparttable}
\usepackage{array}
\newcolumntype{L}[1]{>{\raggedright\arraybackslash}p{#1}}
\DeclareMathOperator{\Tr}{Tr}
\usepackage{color}

\begin{document}

\title{Connection between partial pressure, volatility, and the Soret effect elucidated using simulations of non-ideal supercritical fluid mixtures}

\author[ucfaddress] {Brandon Fields}
\author[ucfaddress,AMPACaddress,REACT]{Patrick K. Schelling\corref{mycorrespondingauthor}}
\cortext[mycorrespondingauthor]{Corresponding author.}
\ead{patrick.schelling@ucf.edu}
\address[ucfaddress]{Department of Physics, University of Central Florida, Orlando, FL 32816-2385, USA}
\address[AMPACaddress]{Advanced Materials Processing and Analysis Center, University of Central Florida, Orlando, FL 32816-2385, USA}
\address[REACT] {Renewable Energy and Chemical Transformations (REACT) Cluster, University of Central Florida, Orlando, FL 32816-2385, USA}

\begin{abstract}
Building on recent simulation work, it is demonstrated using molecular-dynamics (MD) simulations of two-component fluid mixtures that the chemical contribution to the Soret effect in two-component non-ideal fluid mixtures arises due to differences in how the partial pressures of the components respond to temperature and density gradients. Further insight is obtained by reviewing the connection between activity and deviations from Raoult's law in the measurement of the vapor pressure of a liquid mixture. A new parameter $\gamma_{s}^{S}$, defined in a manner similar to the activity coefficient, is used to characterize differences deviations from ``ideal'' behavior.  It is then shown that the difference $\gamma_{2}^{S}-\gamma_{1}^{S}$ is predictive of the
sign of the Soret coefficient and is  correlated to its magnitude. We hence connect the Soret effect to the relative volatility of the components of a fluid mixture, with the more volatile component enriched in the low-density, high-temperature region, and the less volatile component enriched in the high-density, low-temperature region.  Because $\gamma_{s}^{S}$ is closely connected to the activity coefficient, this suggests the possibility that measurement of partial vapor pressures might be used to indirectly determine the Soret coefficient. It is proposed that the insight obtained here is quite general and should be applicable to a wide range of materials systems. An attempt is made to understand how these results might apply to other materials systems including interstitials in solids and multicomponent solids with interdiffusion occurring via a vacancy mechanism.
 \end{abstract}


\maketitle


The Soret effect reflects the tendency of segregation to occur in multicomponent system with an
imposed temperature gradient. In a two-component fluid mixture, the Soret coefficient $S_{T,1}$ is generally the quantity used to quantify the effect in terms of composition and temperature gradients using\cite{Reith_2000},
\begin{equation}
S_{T,1} = -{1 \over x_{1}(1-x_{1})} {\nabla x_{1} \over \nabla T}
 \text{  ,}
\end{equation}
in which $T$ is temperature and $x_{1}$ is the local concentration (mole fraction) of component 1. The sign of $S_{T,1}$ determines the tendency of component 1 to be segregated in the high- or low-temperature regions. Specifically, $S_{T,1}>0$ indicates component 1 will have an enhanced concentration in the low-temperature region, whereas 
$S_{T,1}<0$ indicates that component 1 will have an enhanced concentration in the high-temperature region. In the picture considered in this paper, the system studied is ``open'' with respect to heat inputs, but ``closed'' with respect to particle exchanges with an external reservoir. Specifically, we envision a system where heat is added and removed in local regions leading to a steady-state heat current. Once steady state in achieved, the mass currents are zero corresponding to no particle exchanges with an external reservoir.
 
It was first shown experimentally in benzene-cyclohexane mixtures that the so-called ``chemical'' contribution yields a concentration-dependent $S_{T,1}$\cite{Debuschewitz:2001aa}. Using Lennard-Jones potentials,
Artola and Rousseau\cite{Artola:2007aa} were the first to provide clear molecular-dynamics (MD) simulation results showing how the chemical contribution leads to dependence of $S_{T,1}$ on concentration $x_{1}$. It was shown that $S_{T,1}$ can change sign with varying concentration. It was also shown that the difference in partial enthalpy between the two species is closely connected to the sign correlated with the sign of the Soret coefficient. However, their work yielded only minimal mechanistic understanding of the Soret effect, including why partial enthalpy differences should be correlated with $S_{T,1}$. More recent efforts, which will be detailed later, have provided more mechanistic insight, but there still is not a comprehensive theory that can account for all aspects of the Soret effect\cite{Hoang_2022}.

Recently, molecular-dynamics (MD) simulations of binary Lennard-Jones liquids have demonstrated that the Soret coefficient depends on partial pressures of the two components\cite{Schelling_2023}. It was demonstrated that an input heat pulse generates a pressure gradient which results in the propagation of a compression wave\cite{Schelling_2023}. Both components flow away from the high-temperature region, creating low-density liquid in the high-temperature region, and high-density liquid in the low-temperature region. This induced density gradient itself generates a pressure gradient which opposes the pressure gradient caused by the temperature gradient.   Mechanical equilibrium corresponds to the point where the two effects exactly balance. However, because the two components exhibit different sensitivities to density gradients, the process of reaching mechanical equilibrium results in segregation. Specifically, it was shown that the component which is the most sensitive to density gradients exhibits a higher concentration in the high-temperature region, whereas the component which is the least sensitive to density gradients exhibits a higher concentration in the low-temperature region. This insight represents a new perspective on the Soret effect. However, the calculations reported previously\cite{Schelling_2023} were performed only for equimolar mixtures and hence the concentration dependence typically associated with the chemical contribution was not explored.

In this paper, we further elucidate the Soret effect using MD simulation of supercritical fluids comparable to those studied in Ref. \cite{Artola:2007aa} to elucidate the chemical contribution to the Soret effect. The concentration dependence is explained semi-quantitatively using the insight obtained in the previous work\cite{Schelling_2023}. Based on the previous work, two terms are computed which are correlated to both the sign and the magnitude of the Soret effect. In addition, another perspective based on the relative volatility of the two components is explored by defining a new quantity $\gamma_{s}^{S}$ which is related to the activity coefficient. We show that the difference $\gamma_{2}^{S}-\gamma_{1}^{S}$ is strongly correlated with $S_{T,1}$. This analysis suggests, at least for liquids in equilibrium with a vapor phase, a potential connection between measurements of vapor pressure and the Soret effect. An analysis of recent computational results by Gittus and Bresme \cite{Gittus_2023}, where a minimum in $S_{T,1}$ was first demonstrated in Lennard-Jones (LJ) liquid mitures, is also presented. This analysis shows the importance of the composition dependence of the liquid compressibility in determining the magnitude of $S_{T,1}$. Finally, given that the approach here is rather general, we speculate about thermodiffusion in different materials systems, including interstitials in a solid and multi-component solids where interdiffusion occurs via a vacancy mechanism.

\section{Methods}

Molecular-dynamics (MD) simulations of simple LJ liquid mixtures are used to elucidate the Soret effect. Some of the simulation approaches were used in the previous study\cite{Schelling_2023}. Here the main objective is to obtain insight into the previous results obtained by Artola and Rousseau\cite{Artola:2007aa}, and hence some of the details presented here are related to that work.
The components of the liquid interact via the standard LJ pair potential. The pairwise interaction potential between atoms of component $s$ and $s^{\prime}$ is given in the LJ model by,
\begin{equation}
u_{s s^{\prime}}^{0}(r)=-4\epsilon_{s s^{\prime}} \left[\left(\sigma_{s s^{\prime}} \over r \right)^{6} -\left(\sigma_{s s^{\prime}} \over r \right)^{12} \right]
\end{equation}
Species 1 is taken to be Ar, with $\sigma_{11} = 0.3405$nm, ${\epsilon_{11} \over k_{B}} = 120$K, and $m_{1}=39.948$u. Real physical units are defined 
primarily to obtain a physical timescale and a value for $S_{T,1}$ in K$^{-1}$ so that comparison to Ref. \cite{Artola:2007aa} can be made. Quantities related to energy and pressure are all reported in reduced units. For completeness, the value of the Soret coefficient is also given in reduced units. As in the previous work \cite{Artola:2007aa}, the size and mass parameters for species 2 were taken to be equal to that of species 1. The energy parameter $\epsilon_{22}$ was allowed to vary, with the difference characterized by $\psi_{\epsilon}={\epsilon_{22} \over \epsilon_{11}}$. Cross interactions were specified by the modified Lorenz-Berthelot mixing rule used previously\cite{Artola:2007aa}
\begin{equation}
\epsilon_{12}=\epsilon_{21} = k_{12} \sqrt{\epsilon_{11}\epsilon_{22}}
\end{equation}
where $k_{12}$ is a parameter that characterizes the bonding strength between atoms of component 1 and component 2.

To ensure computational efficiency, a cutoff of $r_{c}=3 \sigma_{11}=3\sigma_{22}$ was used. For atom separations $r>r_{c}$, the interaction was taken to be zero. To ensure no discontinuities in the potential or forces existed, the potentials were smoothed by calculating the interactions from the effective potential,
\begin{equation}
u_{ss^{\prime}}(r)=u_{ss^{\prime}}^{0}(r)-u_{ss^{\prime}}^{0}(r=r_{c})-(r-r_{c})\left[du^{0}_{ss^{\prime}} \over dr\right]_{r=r_{c}} .
\end{equation}
Because the cutoff $r_{c}=3\sigma_{11}$ is quite large, the effect of smoothing is minimal.  After smoothing, the leading discontinuities correspond to second-order derivatives of the potential energy function.

To systematically choose state points for different compositions, the same approached used in Ref. \cite{Artola:2007aa} based on the one-fluid approximation was used \cite{GALLIERO200620}. For two component fluids, $\epsilon_{x}$ which characterizes the interactions, was defined using,
\begin{equation}
\epsilon_{x} = x_{1}^{2} \epsilon_{11} \left[1 + \left({1-x_{1} \over x_{1} }\right)^{2} \psi_{\epsilon} +
{2x_{1}(1-x_{1}) \over x_{1}^{2}} k_{12} \sqrt{\psi_{\epsilon}}  \right] \text{ ,}
\end{equation}
in which $x_{s}$ represents the mole fraction of species $s$, with the requirement for a two-component fluid $x_{1}+x_{2}=1$. 
Following the previous work \cite{Artola:2007aa}, calculations were performed for a super-critical fluid state near $T=1.5T_{c}$ and $\rho = 2.0 \rho_{c}$,  where $T_{c}$ and $\rho_{c}$ represent the critical temperature and density of the mixture. In reduced units, and using the one-fluid approximation value $\epsilon_{x}$, the critical temperature in reduced units is $T_{c}^{*} = 1.316 = {k_{B}T \over \epsilon_{x}}$, and the critical density is $\rho_{c}^{*} =0.304= {N \sigma_{11}^{3} \over V}$ \cite{Johnson_1992,Johnson_1993}.
These conditions were chosen in each case for the equimolar mixture $x_{1}={1 \over 2}$ and the time-averaged pressure was computed. For other compositions, the system volume was varied to obtain the same time-averaged pressure as obtained in the $x_{1}={1 \over 2}$ case. 

To compute partial pressures, the  stress tensor $\bm{\pi}$ is computed using\cite{Parrinello_1981},
\begin{equation}
\boldsymbol{\pi} \Omega  = \sum_{s=1}^{2}\sum_{i=1}^{N_{s}} m_{s} \langle \bm{v}_{si} \bm{v}_{si} \rangle + {1 \over 2}\sum_{s=1}^{2}\sum_{i=1}^{N_{s}}  \sum_{s^{\prime}j\ne s i} \langle \bm{F}_{si,s^{\prime} j} \bm{r}_{si,s^{\prime} j}  \rangle  \text{  ,}
\end{equation}
in which $\bm{v}_{si}$ is a velocity vector of atom $i$ of component $s$,  $\bm{F}_{si,s^{\prime} j} $ 
is the force on atom $i$ of component $s$ due to its interaction with atom $j$ of species $s^{\prime}$, and
 $\bm{r}_{si,s^{\prime} j}=\bm{r}_{si} - \bm{r}_{s^{\prime} j}$ is the relative coordinate vector between the interacting particles. The quantity $\Omega$ is the volume of the simulation supercell.
The total pressure is determined by the trace of the quantity above, namely $p={1 \over 3} \Tr{\left[ \boldsymbol{\pi} \right]}$. To define partial pressures, summation occurs only over one species. The partial stress tensor $\bm{\pi}^{(s)}$ is defined accordingly,
\begin{equation}
\bm{\pi}^{(s)}\Omega  = \sum_{i=1}^{N_{s}} m_{s} \langle \bm{v}_{si} \bm{v}_{si} \rangle + {1 \over 2}\sum_{i=1}^{N_{s}}\sum_{s^{\prime}j\ne s i} \langle \bm{F}_{si,s^{\prime}j} \bm{r}_{si,s^{\prime}j}  \rangle .
\end{equation} 
Then the partial pressures are given by the trace $p_{s}= {1 \over 3} \Tr{\left[ \boldsymbol{\pi}^{(s)} \right]}$.
For a two component system $s=1,2$ then, the relationship between the total pressure and the partial pressures is
\begin{equation}
p = p_{1} + p_{2} .
\end{equation}

It should be noted that the partial pressure as defined above directly relates to mechanical forces and a hydrodynamic description of the fluid.
Several other models of the Soret effect \cite{Montel2019,Shapiro_2020} relate more closely to fugacity which is distinct from partial pressure.
Specifically, fugacity is more directly connected to the chemical potential $\mu_{i}$ of component $i$. For example, $\mu_{i}$ in the vapor mixture is given by the expression,
\begin{equation}
\mu_{i}=\mu^{\ominus}_{i,vap} +RT \ln{f_{i} \over p^{\ominus}} \text{  ,}
\end{equation}
in which $R$ is the gas constant, $f_{i}$ is the fugacity, and $\mu^{\ominus}_{i,vap}$ is a reference chemical  potential at fugacity $f_{i}=p^{\ominus}$. In an ideal gas, $f_{i}$ is equal to the partial pressure $p_{i}$. In a real gas with attractive interactions between gas molecules, the fugacity is lower than the partial pressure, $f_{i}<p_{i}$. 
An analogous expression can be written for a liquid phase, and equilibrium between liquid and vapor phases corresponds to equal values of the fugacity in each phase. Starting from chemical potential gradients within a linear non-equilibrium formalism is a standard approach for studying thermodiffusion, and hence the connection to 
fugacity naturally develops. However, in this paper, we specifically determine a connection between equilibrium partial pressures and the Soret effect. While partial pressure and fugacity are closely related, we have not yet established connections to previous theories that base the concept of fugacity.

The microscopic definition of the local particle flux densities for the two components are given by,
\begin{equation}
\bm{J}_{s}(\bm{r},t) = \sum_{j=1}^{N_{s}} \bm{v}_{sj}(t) \delta^{(3)} (\bm{r}-\bm{r}_{sj})
\text{  .}
\end{equation}
Here $N_{s}$ represents the total number of particles of component $s$. 
The particle densities for the two species are defined by,
\begin{equation} \label{dens}
n_{s}(\bm{r}) = \sum_{j=1}^{N_{s}} \delta^{(3)} (\bm{r}-\bm{r}_{sj}).
\end{equation}
The Fourier transforms of the particle current densities are represented by,
\begin{equation} \label{js}
\tilde{J}_{s}(\bm{k},t) ={1 \over \Omega}  \sum_{j=1}^{N_{s}} { \bm{v}_{sj}(t) \cdot \bm{k} \over |\bm{k}|} e^{-i \bm{k} \cdot \bm{r}_{sj}(t)} \text{  ,}
\end{equation}
in which we use that longitudinal currents are relevant for the evolution of density and concentration gradients. The component densities are also represented in reciprocal space with,
\begin{equation} \label{js}
\tilde{n}_{s}(\bm{k},t) ={1 \over \Omega}  \sum_{j=1}^{N_{s}} e^{-i \bm{k} \cdot \bm{r}_{sj}(t)} \text{  .}
\end{equation}
The vectors $\bm{k}$ are the reciprocal lattice vectors for the simulation supercell with periodic-boundary conditions applied in each dimension.

The liquid was generated by melting an initial randomized structure with atoms of both sites situated randomly at FCC lattice sites. The systems contained $4000$ atoms in a cubic simulation cell with different numbers $N_{1}$ and $N_{2}$ for the two types.  The equilibrium concentration of component 1 is given by $x_{1}={N_{1} \over N_{1}+N_{2}}$.  The MD integration timestep was $\Delta t=2.15$fs, with a reduced timestep $\Delta t^{*} = 0.001$. This was verified to result in good energy conservation. Most quantities below, including temperature, pressure, and density, are reported in reduced units. Time-dependent results are displayed with times in picoseconds. Also, the Soret coefficient $S_{T,1}$ is reported in dimensions $K^{-1}$ to compare with previous results\cite{Artola:2007aa}. However, the reduced value $S_{T,1}^{*} = S_{T,1} \epsilon_{11} k_{B}^{-1} $ is also reported for completeness. In the Supplementary file, data related to the simulation conditions, including density, temperature, and pressure can be obtained. 

There are two main types of calculations reported here. In the first, the partial pressures of the two components are computed after a long equilibrium simulation in the constant volume, particle number, and temperature ensemble. Equilibrium averaging included $1.075$ns of simulation time to ensure converged values of the partial pressures. Calculations were performed to determine the dependence of partial pressures on temperature and densities of the two components. Temperature dependence was determined by computation of $p_{1}$ and $p_{2}$ for three different temperatures. Then partial derivatives with respect to temperature were determined numerically. Density dependence of the partial pressures $p_{1}$ and $p_{2}$ at a given concentration $x_{1}$ and reduced temperature $T^{*}$ were obtained by removing or adding 20 atoms of one component. Partial derivatives of $p_{1}$ and $p_{2}$ with respect to densities were computed numerically from the results. When adding 20 atoms, the extra atoms were placed at random interstitial sites of the FCC lattice followed by the usual melting process.

In the second type of calculation, a steady-state external heat perturbation is applied to generate a temperature and composition gradient. As in the previous work\cite{Schelling_2023}, an sinusoidal heat source was implemented by velocity rescaling at each MD timestep,
\begin{equation}\label{pert}
\bm{v}_{sj}(t) \rightarrow \bm{v}_{sj}(t) \sqrt{1 + b \cos{\left(k z_{sj}\right)}} \text{ ,}
\end{equation}
in which $z_{sj}$ is the z Cartesian coordinate of atom $j$ within component $s=1,2$. The quantity $b=2 \times 10^{-5}$ was used in the calculations.
The value of $b$ determines the strength of the perturbation, and $k$ was chosen to be $k={2 \pi \over L}$, where $L$ is the length of the simulation supercell along the $z$ axis. To ensure that this heat source did not generate translation of the system by imparting a net momentum, any deviations from zero total momentum for the system were cancelled by subtraction of a small velocity component applied uniformly to all atoms in the system. After sufficient time, the temperature and concentration profiles are found to achieve a steady state, and then the Soret coefficient $S_{T,1}$ can be approximately determined.

For steady-state calculations, the Soret coefficient is computed using,
\begin{equation} \label{soret}
S_{T,1}(t) = -{3k_{B} \over 2} { \left(\tilde{n}_{1}(k,t) \middle/ x_{1}\right) -\left(\tilde{n}_{2}(k,t) \middle/(1-x_{1}) \right) \over \tilde{E}_{K}(k,t)
-{3 \over 2}\tilde{n}(k,t) k_{B}T_{0}} \text{ ,}
\end{equation}
in which $\tilde{n}_{s}(k,t)$ are the densities of the components $s$ in reciprocal space, $x_{1}$ is the average (bulk) concentration, $T_{0}$ is the average system temperature, $\tilde{n}=\tilde{n}_{1}+\tilde{n}_{2}$, and $\tilde{E}_{K}$ is the kinetic-energy density in reciprocal space given by the equation,
\begin{equation}
\tilde{E}_{K}(k,t) = {1 \over \Omega} \sum_{s=1}^{2} \sum_{j=1}^{N_{s}} {1 \over 2}
m_{s}\left[\bm{v}_{sj}(t)\cdot \bm{v}_{sj}(t)\right] e^{-ikz_{sj}}
\end{equation}
Equation \ref{soret} is related to the equation used previously in Ref.\cite{Schelling_2023}, but this is more general for different concentrations and also includes terms that were previously neglected. It is expected that $S_{T,1}(t)$ computed using Eq. \ref{soret} will eventually converge to a steady-state value once $t$ is large enough for the temperature and concentration profiles to stabilize. After reaching steady state, $S_{T,1}$ is obtained by time averaging the quantity in Eq. \ref{soret}.

It is recognized that the use of a sinusoidal heat source has not been standard in nonequilibrium MD simulations of the Soret effect nor in calculations of thermal conductivity. The more standard approach is to use separated hot and cold reservoirs. However, in the previous work, we have found the use of sinusoidal sources particularly useful for elucidating transients\cite{Schelling_2023}. Moreover, we have established the general approach in previous thermal conductivity studies of LJ solids\cite{Fernando_2020} and Fermi-Pasta-Ulam chains\cite{Bohm:2022aa}.  Specifically, in Ref. \cite{Fernando_2020}, we were able to demonstrate that sinusoidal heat sources were able to extrapolate to the bulk thermal conductivity obtained using Green-Kubo calculations in the limit that the wavelength of the source approaches infinity. However, in view of the fact that this approach has not been standard in MD simulations of the Soret effect, we include comparisons to a more standard approach in the next section. Specifically we will perform calculations of the calculation of the Soret coefficient $S_{T,1}$ from time-averaged temperature and concentration gradients for separated hot and cold reservoirs. We will also demonstrate that the reported calculations are represent steady-state averages and that the response is well within the linear regime.

\section{Results}

We first discuss calculations of the relevant partial thermodynamic quantities which will be related to results for $S_{T,1}$. Relevant details
for the simulation conditions are reported in the Supplementary document. 
The results reported in Table \ref{table1} correspond to simulations with $k_{12}=1.35$ and 
$\psi_{\epsilon}=1.0$. Results are only shown for $x_{1}$ between $0.10$ and $0.50$. In this case, fluids with concentration $x_{1}$ and $1-x_{1}$ are equivalent, but with components $1$ and $2$ swapped. Hence, calculation of results for $x_{1}>0.50$ would be redundant.

Several trends emerge from Table \ref{table1}. First, as expected, the partial pressures $p_{1}$ and $p_{2}$ depend in a rather obvious way on $x_{1}$, with the total $p_{1}+p_{2}$ being very
close for each different concentration $x_{1}$. The computed partial derivatives also depend very strongly on $x_{1}$. To determine the relationship of the quantities in Table \ref{table1} and computed values of $S_{T,1}$, however, requires additional insight. It should be noted that, with the exception of the equimolar mixture $x_{1}=0.50$ which has $S_{T,1}=0$, each composition here was found in Ref. \cite{Artola:2007aa} to have $S_{T,1}>0$, indicating the tendency of enhance concentration of component 1 in the low-temperature region.  For $x_{1} >0.50$, $S_{T,1}<0$ was obtained, indicating that component 1 has an enhanced concentration in the high-temperature region.

In Table \ref{table2}, the same quantities are reported for $\psi_{\epsilon}=1.80$ and $k_{12}=1.35$. In Ref. \cite{Artola:2007aa}, there is a strong tendency for $S_{T,1}<0$, with the possible exception of concentrations $x_{1}\lesssim 0.20$. This indicates that choosing a value for $\psi_{\epsilon}$ greater than 1 leads to a trend for increased concentrations of component 1 in the high-temperature region. Later we report our own calculation of the Soret coefficients. The results in Ref.\cite{Artola:2007aa} indicate very small values of $S_{T,1}$ near $x_{1}=0.20$ with potentially enough statistical error to make the determination of a concentration-dependent sign change very difficult. As detailed later in this work, our results indicate $S_{T,1}<0$ for all the simulated concentrations $x_{1}$ in apparent contradiction to Ref.\cite{Artola:2007aa}. In the Supplementary document, the simulation parameters, including density and temperature, for these calculations are reported. This should enable reproduction of our results and clarification of any deviations from Ref. \cite{Artola:2007aa}. 

\begin{table}
\begin{center}
\caption{Tabulated results for quantities relevant to generalized thermodynamic forces for simulations
with $k_{12}=1.35$ and $\psi_{\epsilon}=1.0$. Results are obtained for different values of concentration $x_{1}$ of component $1$. Quantities everywhere are given in reduced units. Analyzing the statistical distributions of partial pressures, the error bars for the partial derivatives here are at most $\pm 0.05$.
}
\begin{tabular} {|c|c|c|c|c|c|c|c|c|}
\hline
$x_{1}$ & $ p_{1}$ & $p_{2} $ & $ \left({\partial p_{1} \over \partial T^{*}}\right)_{n_1,n_{2}}$ &   $ \left({\partial p_{2} \over \partial T^{*}}\right)_{n_1,n_{2}}$   & $\left({\partial p_{1} \over \partial n_{1}}\right)_{T^{*},n_{2}}$   & $\left({\partial p_{2} \over \partial n_{1}}\right)_{T^{*},n_{2}}$ &  $\left({\partial p_{1} \over \partial n_{2}}\right)_{T^{*},n_{1}}$ &   $\left({\partial p_{2} \over \partial n_{2}}\right)_{T^{*},n_{1}}$ \\\hline 
$0.10$ & 0.208 & 2.339 & 0.221  & 1.913 & 4.128  & 7.183  & 0.775 & 11.814   \\\hline
$0.20$ & 0.459 & 2.102 & 0.446  & 1.644  & 5.572 &  6.405 & 1.566 &  11.470   \\\hline
$0.30$& 0.712 & 1.839  & 0.615  & 1.419 & 6.572 & 5.523 & 2.376  & 10.500 \\\hline
$0.40$& 0.987 & 1.554  & 0.798  & 1.191  & 7.707  & 4.524 & 3.266  & 9.644   \\\hline
$0.50$& 1.273 & 1.273  & 0.988  & 0.988  & 8.335 & 4.153  &  4.153 & 8.335 \\\hline
\end{tabular}
\label{table1}
\end{center}
\end{table}

\begin{table}
\begin{center}
\caption{Tabulated results for quantities relevant to generalized thermodynamic forces for simulations
with $k_{12}=1.35$ and $\psi_{\epsilon}=1.80$. Results are obtained for different values of concentration $x_{1}$ of component $1$. Quantities everywhere are given in reduced units. Analyzing the statistical distributions of partial pressures, the error bars for the partial derivatives here are at most $\pm 0.05$.
}
\begin{tabular} {|c|c|c|c|c|c|c|c|c|}
\hline
$x_{1}$ & $ p_{1}$ & $p_{2} $ & $ \left({\partial p_{1} \over \partial T^{*}}\right)_{n_1,n_{2}}$ &   $ \left({\partial p_{2} \over \partial T^{*}}\right)_{n_1,n_{2}}$   & $\left({\partial p_{1} \over \partial n_{1}}\right)_{T^{*},n_{2}}$   & $\left({\partial p_{2} \over \partial n_{1}}\right)_{T^{*},n_{2}}$ &  $\left({\partial p_{1} \over \partial n_{2}}\right)_{T^{*},n_{1}}$ &   $\left({\partial p_{2} \over \partial n_{2}}\right)_{T^{*},n_{1}}$ \\\hline 
$0.10$ & 0.354 & 3.103 &  0.180 & 1.665 &  7.323 & 10.006 & 1.026 & 16.152   \\\hline
$0.20$ & 0.725 & 2.724  & 0.356 & 1.471  &8.811 & 8.683  & 1.983  & 14.705     \\\hline
$0.30$ & 1.114 & 2.353  &  0.552 & 1.329 &10.026 & 7.342 & 3.185 & 13.285 \\\hline
$0.40$ & 1.504 & 1.958  & 0.751 & 1.173  & 11.554 & 5.998  & 4.090  & 12.509   \\\hline
$0.50$ & 1.888 & 1.562  & 0.998 & 0.982 & 12.791 & 5.340 & 5.682  & 11.095  \\\hline
$0.60$ & 2.271 & 1.187  & 1.194 & 0.842 & 14.188  & 4.085  & 6.140 & 9.043  \\\hline
$0.70$ & 2.643 & 0.821   & 1.507& 0.677 & 15.337  & 3.408 & 7.294  & 7.525  \\\hline
$0.80$ & 2.979 & 0.485  & 1.828 & 0.451 &  16.272 & 2.053  & 8.524   & 5.851  \\\hline
$0.90$ & 3.249 & 0.202  & 2.257 & 0.266  & 17.076 & 0.981  & 9.680 & 4.236  \\\hline
\end{tabular}
\label{table2}
\end{center}
\end{table}

To connect the results for $S_{T,1}$ repored in Ref. \cite{Artola:2007aa} and those obtained in our own $S_{T,1}$ calculations to the numerical data in Tables \ref{table1}-\ref{table2}, we build on the insight developed in the previous paper on the Soret effect in LJ fluids\cite{Schelling_2023}. It was found there that the dominant effect is the difference in the responses of the two fluid components to changes in density. A heat pulse generates a pressure gradient. The response of the system is the generation of a damped compression wave, which drives both components away from the high-temperature region and towards the low-temperature region, generating, respectively, a high-density region and a low-density region. The system approaches mechanical equilibrium by balancing the pressure gradients imposed by the heat source and the resulting density gradient. However, because the responses of the two components to density variations differ, the two components become segregated due to gradients in partial pressures.

The basic ideas presented below result in insight, but fall short of a more complete theory that could result in quantitative predictions for $S_{T,1}$. However, the arguments below demonstrate the overall composition-dependent trends and hence the chemical contribution to the Soret effect, and hence seem to be an important step towards a more comprehensive theory. Essentially, the analysis provided here we believe establishes the relevance of the quantities shown in Tables \ref{table1}-\ref{table2} to $S_{T,1}$.

To develop these ideas using the insight in Ref. \cite{Schelling_2023} as a starting point, we must include concentration dependence. In the previous work, only equimolar concentrations $x_{1}=0.5$ were used. We first note that a particle flux density can be written as a density multiplied by a local velocity field. So for component $s$, we can write a local mass flux at time $t$,
\begin{equation}
J_{s}(z,t) = n_{s}(z,t) v_{s}(z,t) \text{ ,}
\end{equation}
in which the particle flux density $J_{s}(z)$, density $n_{s}(z)$, and velocity field $v_{s}(z)$ are only dependent on the Cartesian coordinate $z$ when the a sinusoidal source is applied in such a way as to generate gradients directed along the Cartesian $z$-axis.
This is a more macroscopic description in contrast to the microscopic definitions introduced above. Because we are interested in understanding the driving force which develops the concentration gradient, here we focus on the relative acceleration ${\partial v_{1}\over \partial t} - {\partial v_{2} \over \partial t}$. Concentration gradients then emerge due to differences between $v_{1}$ and $v_{2}$. Due to the application of a sinusoidal heating source, we work in Fourier space with $k={2 \pi \over L}$ and then examine the quantity,
\begin{equation} \label{v12}
{\partial \tilde{v}_{1}(k,t)\over \partial t} - {\partial\tilde{v}_{2}(k,t) \over \partial t}
= {\tilde{F}_{1}(k,t) \over m_{1} N_{1}}  -
  {\tilde{F}_{2}(k,t) \over m_{2} N_{2}}  \text{ ,}
\end{equation}
in which $\tilde{F}_{s}(k,t) $ are forces terms evaluated in Fourier space\cite{Schelling_2023}.  This expression indicates the relevant terms for the isotope effect. A discussion of the isotope effect appears in the Conclusions section. For the simulations here, $m_{1}=m_{2}=m$, and hence we have,
\begin{equation}
{\partial \tilde{v}_{1}(k,t)\over \partial t} - {\partial \tilde{v}_{2}(k,t) \over \partial t} 
= {1 \over mN } \left[{\tilde{F}_{1}(k,t) \over x_{1}}  -
  {\tilde{F}_{2}(k,t) \over  1- x_{1}}\right] \text{ ,}
\end{equation}
and hence the Soret effect occurs here only due to differences in the driving forces in Fourier space.
Now using a microscopic description, 
the force terms in Fourier space are obtained at time $t$ from,
\begin{equation}
\tilde{F}_{s}(k,t) = \sum_{j=1}^{N_{s}} F_{sj,z}(t) e^{-ikz_{sj}(t)} \text{ ,}
\end{equation}
and $\tilde{v}_{s}(k,t)$ are the Fourier components of the velocity fields for species $s$,
\begin{equation}
\tilde{v}_{s}(k,t) = \sum_{j=1}^{N_{s}}  v_{sj,z}(t) e^{-ikz_{sj}(t)} \text{ .}
\end{equation}
Following Ref \cite{Schelling_2023}, the forces in Fourier space are due to the temperature and density gradients. Specifically, the force term for each component $s$ is hypothesized to depend on Fourier components $\tilde{T}(k)$ for the temperature profile, and $\tilde{n}_{1}(k) $, $\tilde{n}_{2}(k)$ for the density profiles as, 
\begin{equation} \label{fs}
\tilde{F}_{s}(k) =  -i k \Omega \left[\left({\partial p_{s} \over \partial T} \right)_{n_{1},n_{2}} 
\tilde{T}(k)   +
\left({\partial p_{s} \over \partial n_{1}} \right)_{T,n_{2}}\tilde{n}_{1}(k) +
\left({\partial p_{s} \over \partial n_{2}} \right)_{T,n_{1}} \tilde{n}_{2}(k) \right]
\end{equation}  
The fact that the Fourier space terms are complex are evident from this equation. In fact, since the external heating profile is represented in space by a cosine function (see Eq. 13), the density and temperature profiles take on the same cosine profile. By contrast, the Fourier components of the forces $\tilde{F}_{s}(k)$, since they depend on gradients of the temperature and density, can be represented in real space by a sine function. This is reflected in the $i=\sqrt{-1}$ factor on the right-hand side of Eq. \ref{fs}.

The next approximation used to analyze the response is to assume to zeroth order that the Fourier components of the density functions are given by,
 \begin{equation}
 \tilde{n}_{1}(k,t) \approx x_{1}\tilde{n}(k,t) 
  \end{equation}
   \begin{equation}
 \tilde{n}_{2}(k,t) \approx (1-x_{1})\tilde{n}(k,t)  \text{ .}
 \end{equation}
Here $\tilde{n}(k,t)= \tilde{n}_{1}(k,t) + \tilde{n}_{2}(k,t)  $ represents a Fourier component of the total particle density when driven from equilibrium. This approximation captures the fact that the deviations from equilibrium $\tilde{n}_{1}(k,t)$ and $\tilde{n}_{2}(k,t)$ are primarily determined by the overall density profile via $\tilde{n}(k,t) $ and also the equilibrium concentrations $x_{1}$ and $x_{2}=1-x_{1}$. While concentration gradients are obviously the focus in a study of the Soret effect, we can justify this approximation by the fact that we are considering the early stages of development of the concentration gradient before the concentration gradient is yet to fully evolve to the steady state. Moreover, even after the concentration gradient reaches the steady state, concentration gradients are generally relatively small.

With these assumptions,  the evolution of the concentration gradient can be described by,
\begin{equation}\label{dvdt}
\begin{split}
mN & \left[{\partial \tilde{v}_{1}(k,t)\over  \partial t} - {\partial \tilde{v}_{2}(k,t) \over \partial t} \right] 
\approx -ik \Omega
 \left[\left({1 \over x_{1}} \right)\left({\partial p_{1} \over \partial T} \right)_{n_{1},n_{2}} 
 -  \left({1 \over 1- x_{1}} \right)\left({\partial p_{2} \over \partial T} \right)_{n_{1},n_{2}}\right]
\tilde{T}(k,t) + \\
  &  \left[ \left({\partial p_{1} \over \partial n_{1}} \right)_{T,n_{2}} +
\left({1-x_{1} \over x_{1}}\right)\left({\partial p_{1} \over \partial n_{2}} \right)_{T,n_{1}}-
 \left({x_{1} \over 1-x_{1}}\right)\left({\partial p_{2} \over \partial n_{1}} \right)_{T,n_{2}}-
\left({\partial p_{2} \over \partial n_{2}} \right)_{T,n_{1}}
\right]\tilde{n}(k,t)  
\end{split}
\end{equation}
In developing this equation, we are attempting to elucidate the physics which controls the magnitude and the sign of the Soret effect. The equation above cannot directly
be used to make a quantitative prediction about the exact value of $S_{T,1}$ in a steady-state experiment or simulation. However, Eq. \ref{dvdt} is nevertheless useful for elucidating the physics, and moreover suggests a general direction for the development of a quantitative theory based on equilibrium properties. 

\begin{figure}
\begin{centering}
\includegraphics[width=0.5\textwidth]{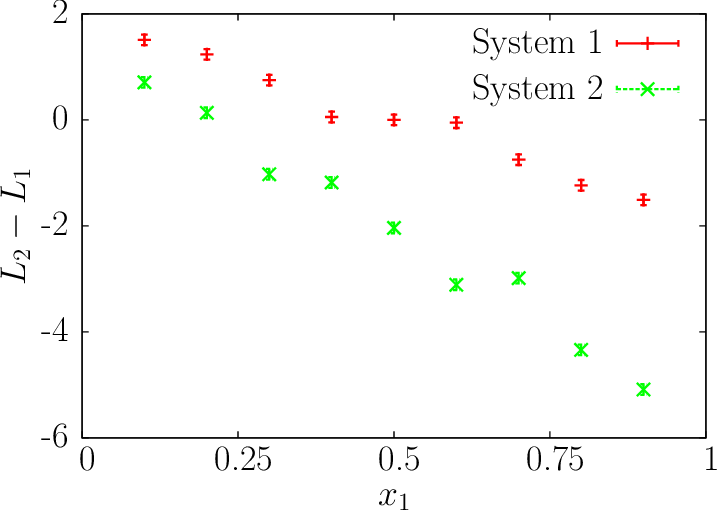} 
\caption{The quantity $L_{2}-L_{1}$ plotted for $k_{12}=1.35$ and $\psi_{\epsilon}=1.0$  plotted for System 1 ($k_{12}=1.35$, $\psi_{\epsilon}=1.0$) and System 2 ($k_{12}=1.35$,$\psi_{\epsilon}=1.80$) simulations as a function of concentration $x_{1}$.
}
\label{fig1}
\end{centering}
\end{figure}

\begin{figure}
\begin{centering}
\includegraphics[width=0.5\textwidth]{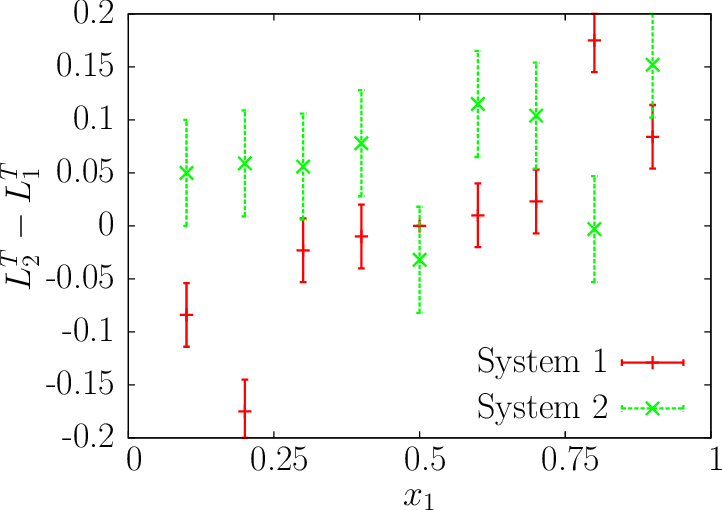} 
\caption{The quantity $L^{T}_{2}-L^{T}_{1}$ plotted for System 1 ($k_{12}=1.35$, $\psi_{\epsilon}=1.0$) and System 2 ($k_{12}=1.35$,$\psi_{\epsilon}=1.80$) simulations as a function of concentration $x_{1}$.
}
\label{fig2}
\end{centering}
\end{figure}

\begin{figure}
\begin{centering}
\includegraphics[width=0.5\textwidth]{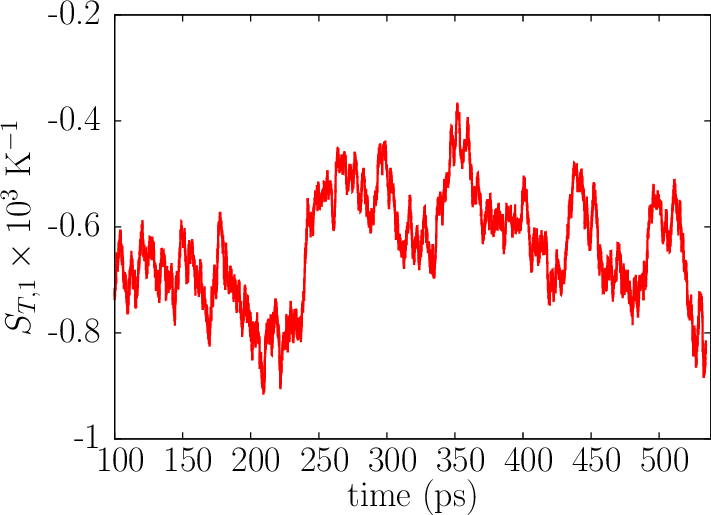} 
\caption{The Soret coefficient $S_{T,1}(t)$ plotted as a function of time $t$ for $k_{12}=1.35$, $\psi_{\epsilon}=1.80$, and concentration $x_{1}=0.30$. The data here is averaged over 1440 independent simulations. Time averaging beyond $200$ps results in the value $S_{T,1}=-0.618 \times 10^{-3}$K $^{-1}$ as shown in Table \ref{table4}.
}
\label{fig3}
\end{centering}
\end{figure}

\begin{figure}
\begin{centering}
\includegraphics[width=0.5\textwidth]{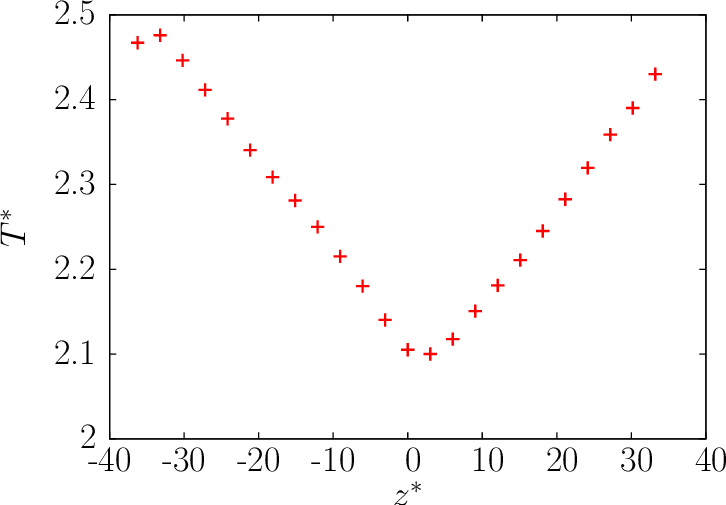} 
\caption{The reduced temperature $T^{*} = {k_{B}T \over \epsilon_{11}}$ plotted as a function of position $z^{*}={z \over \sigma_{11}}$ for NEMD simulations. Results here correspond to bulk concentration $x_{1}=0.90$ with parameters $k_{12}=1.35$ and $\psi_{\epsilon} = 1.80$.
}
\label{fig4}
\end{centering}
\end{figure}

We next analyze separately the relevant terms driving the separation based on density gradients and temperature gradients. First note that temperature gradients tend to drive both components from the high to low temperature regions. This is obvious given the tendency of fluids to expand when heated. It is reflected in the fact that $\left({\partial p_{1} \over \partial T} \right)_{n_{1},n_{2}}>0 $ and $\left({\partial p_{1} \over \partial T} \right)_{n_{1},n_{2}}>0 $. As fluid flows away from the high-temperature region to the low-temperature, the pressure builds up due to the terms related to concentration gradients. This tends to counterbalance the pressure gradients due to the external heat source. This is reflected in the fact that each
partial derivative with respect to density in Eq. \ref{dvdt} is also positive. Given these considerations, we analyze the terms defined below related to density gradients,
\begin{equation}
L_{1} = \left({\partial p_{1} \over \partial n_{1}} \right)_{T,n_{2}} +
\left({1-x_{1} \over x_{1}}\right)\left({\partial p_{1} \over \partial n_{2}} \right)_{T,n_{1}}
\end{equation}
\begin{equation}
L_{2} =  \left({x_{1} \over 1-x_{1}}\right) \left({\partial p_{2} \over \partial n_{1}} \right)_{T,n_{2}}+
\left({\partial p_{2} \over \partial n_{2}} \right)_{T,n_{1}}
\end{equation}
and those due to temperature gradients,
\begin{equation}
L_{1}^{T} = \left({1 \over x_{1}} \right)\left({\partial p_{1} \over \partial T} \right)_{n_{1},n_{2}}
\end{equation}
\begin{equation}
L_{2}^{T} = \left({1 \over 1- x_{1}} \right)\left({\partial p_{2} \over \partial T} \right)_{n_{1},n_{2}}
\end{equation}
The central hypothesis of this paper is that the Soret effect, namely the sign and magnitude of $S_{T,1}$, is determined by the relative difference $L_{2}-L_{1}$, with difference $L_{2}^{T}-L_{1}^{T} $ playing a secondary role. In our previous work, this conclusion was established only for equimolar mixtures $x_{1}=0.5$, and the quantities above were also not explicitly defined\cite{Schelling_2023}. 
In the Supplementary document, the results for $L_{1}$, $L_{2}$, $L_{1}^{T}$, and $L_{2}^{T}$ are reported in tables. Below, the results for $L_{2}-L_{1}$ and $L_{2}^{T}-L_{1}^{T}$ are presented in Figures \ref{fig1}-\ref{fig4} along with discussion.

The analysis begins with  the dependence of $L_{2}-L_{1}$ on concentration $x_{1}$ shown in Fig. \ref{fig1} for both of the systems based on the data in Tables \ref{table1}-\ref{table2}. This quantity is correlated with the relative tendency of the two components to resist accumulation in the low-temperature dense region. Specifically, when $L_{2}-L_{1}>0$, component 2 is more resistant to accumulation in the low-temperature, high-density region.  Hence, the positive values of $L_{2}-L_{1}$ for $x_{1}<0.5$ are consistent with the calculations reported in Ref. \cite{Artola:2007aa} which show clearly that $S_{T,1}>0$ for $x_{1}<0.5$. For $x_{1}>0.5$, $L_{2}-L_{1}<0$, and then $S_{T,1}$ becomes negative\cite{Artola:2007aa}.  For the equimolar case $x_{1}=0.5$, the two components are identical and hence $S_{T,1}=0$ is expected.
The dependence of $L_{2}-L_{1}$ on concentration $x_{1}$ is also shown in Fig. \ref{fig1} for the system with $k_{12}=1.35$ and $\psi_{\epsilon}=1.80$. In this system, $L_{2}-L_{1}<0$ with the exception of the cases with $x_{1} \leq 0.20$. This is consistent with the results in Ref. \cite{Artola:2007aa}, which show $S_{T,1}<0$ with the possible exception of very low concentrations $x_{1} \leq 0.20$. Specifically, for most of the compositions, the results in Ref. \cite{Artola:2007aa} demonstrate accumulation of component $s=2$ in the high-density, low-temperature region. 

Next the quantity $L_{2}^{T}-L_{1}^{T}$ is analyzed to determine the importances of different driving forces that arise directly from the temperature gradient. The results are shown in Fig. \ref{fig2} for $k_{12}=1.35$ and $\psi_{\epsilon}=1.0$, and also for $k_{12}=1.35$ and $\psi_{\epsilon}=1.80$. The data is also given in tables in the Supplementary document. These results show relatively small values which show less dramatic  and systematic dependence on composition $x_{1}$. It is possible some of the scatter of the data is due to small differences in total pressure across the range of $x_{1}$ values. However, the observation that the terms directly related to temperature gradients appear to be less important for explaining the sign and the magnitude of $S_{T,1}$ made in our previous work \cite{Schelling_2023} appears to be true here as well.

In summary, the values of $L_{2}-L_{1}$ appear to be the most important for determining the sign and magnitude of $S_{T,1}$ values reported in Ref. \cite{Artola:2007aa}. However, for more certainty on this point, and especially to explore concentrations $x_{1}$ where $S_{T,1}$ is small and, as reported in Ref. \cite{Artola:2007aa}, undergoes a concentration-dependent change in sign, several independent calculations of $S_{T,1}$ were made for the system with $k_{12}=1.35$ and $\psi_{\epsilon}=1.80$. 

The $S_{T,1}(t)$ calculations were done according to Eq. \ref{soret}. Simulations required less than $200$ps of simulation time to reach steady-state conditions. The results for $S_{T,1}(t)$ were obtained from averaging a large number of independent simulations. Then the reported Soret coefficient $S_{T,1}$ obtained by time-averaging over $330$ps of data. In Fig. \ref{fig3}, the quantity $S_{T,1}(t)$ from Eq. \ref{soret} is plotted as a function of time for $k_{12}=1.35$ and $\psi_{\epsilon}=1.80$ and concentration $x_{1}=0.30$. For this case 1440 independent simulations were averaged. While statistical noise is evident, $S_{T,1}(t)$ appears to reach a steady-state before $200ps$ when time averaging begins.

The results are shown in Table \ref{table3} for $k_{12}=1.35$ and $\psi_{\epsilon}=1.0$ and Table \ref{table4} for $k_{12}=1.35$ and $\psi_{\epsilon}=1.80$. The results in Table \ref{table3} appear to be in agreement with Ref. \cite{Artola:2007aa} within the statistical error. However,
the results in Table \ref{table4} suggest our computed values for $S_{T,1}$ are larger in magnitude by approximately a factor of $2$, even though the qualitative trend is very similar. For this particular case, in our calculations, $S_{T,1}<0$ is obtained for all concentrations. By contrast, the results in Ref. \cite{Artola:2007aa} report a sign change with positive values for $S_{T,1}$ for $x_{1} \leq 0.20$. However, the decrease in magnitude of $S_{T,1}$ as $x_{1}$ decreases is qualitatively similar to the results in Ref. \cite{Artola:2007aa}. While we are unsure of the reasons for the disagreement with Ref. \cite{Artola:2007aa}, the detailed conditions for the calculations reported here are given in the Supplementary document which should enable reproduction. 

\begin{figure}
\begin{centering}
\includegraphics[width=0.5\textwidth]{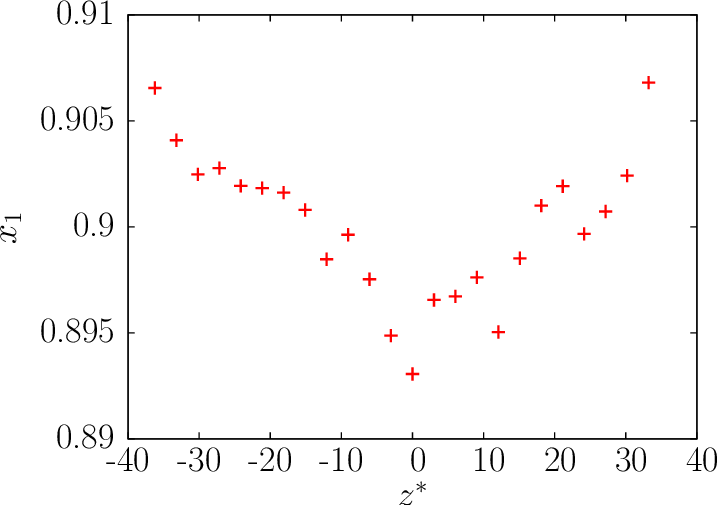} 
\caption{The concentration $x_{1}$ plotted as a function of position $z^{*}={z \over \sigma_{11}}$ for NEMD simulations. Results here correspond to bulk concentration $x_{1}=0.90$ with parameters $k_{12}=1.35$ and $\psi_{\epsilon} = 1.80$.
}
\label{fig5}
\end{centering}
\end{figure}

To further establish the method of using sinusoidal perturbations, we have also computed $S_{T,1}$ for only $k_{12}=1.35$, $\psi_{\epsilon}=1.80$, and $x_{1}=0.90$ using a more standard non-equilibrium MD (NEMD) approach with a heat source and sink. We specifically use the velocity rescaling algorithm due to Jund and Jullien\cite{Jund:1999aa}. For a system with $4000$ total atoms, $30$ calculations of $S_{T,1}$ were performed, each with an independent starting configuration. The simulation supercell for these calculations was longer in the transport direction, with $L_{z} \approx 72.45 \sigma_{11}$. The thermostatted regions had length equal to $0.10L_{z}$. Each of the 30 independent simulations included $1076$ps of data, with time averaged temperature profiles computed over the last $215$ps. In Figs. \ref{fig4}-\ref{fig5}, the temperature and concentration profiles are shown. While the concentration profile is somewhat noisy, it is still possible to calculate $S_{T,1}$.  The computed value $S_{T,1}=(-2.5 \pm 0.4)\times 10^{-3}$ K$^{-1}$ was obtained. Within the statistical error then, the NEMD approach described here agrees with the single data point  $S_{T,1}=(-2.800 \pm 0.120)\times 10^{-3}$ K$^{-1}$ in Table \ref{table4} obtained using sinusoidal perturbations. This suggests the data in Tables \ref{table3}-\ref{table4} are reliable.

We also explored whether the simulations for the perturbation parameter $b=2 \times 10^{-5}$ correspond to a linear-response regime. As expected, smaller values of $b$ result in substantially more statistical error. Specifically we computed results for $k_{12}=1.35$, $\psi_{\epsilon}=1.80$, and composition $x_{1}=0.30$ for 5 different values of $b$. The results are shown in Table \ref{table5}. It is seen that for smaller values $b<2 \times 10^{-5}$, results are consistent although the statistical error becomes quite substantial. 
For $b=4 \times 10^{-5}$, $S_{T,1}$ is notably smaller indicating some nonlinearity in the response. However, for $b=1 \times 10^{-5}$ and $b=2 \times 10^{-5}$, the results are quite close and hence we conclude $b=2 \times 10^{-5}$ is essentially in a linear-response regime.

While these results seem to establish our approach, there is a need for more in-depth methodological study. It should be noted that in Ref. \cite{Fernando_2020}, the sinusoidal perturbation approach was applied to compute thermal transport in LJ solids. There it was demonstrated that the effective thermal conductivity could be extrapolated to the bulk Green-Kubo result in the limit of infinite system size. Finite-size effects have also recently been shown to be relevant in the calculation of $S_{T,1}$ using NEMD simulations\cite{Gittus_2023}. Finite-size effects in MD calculations of diffusion have also been recently reported\cite{Jamali_2018}. This deserves future scrutiny in the context of the rather small sizes used for the sinusoidal perturbations. It may be that finite-size dependence is connected to the question of linear response. However, it is reasonable to expect that values of $S_{T,1}$ computed here are sufficient to establish trends with respect interaction parameters and concentrations. 


\begin{table}
\begin{center}
\caption{Values of $S_{T,1}$ obtained from steady-state calculations for the system
with $k_{12}=1.35$ and $\psi_{\epsilon}=1.0$ as a function of composition $x_{1}$. Results were obtained from $720$ independent simulations each with $S_{T,1}$ averaged over $330$ps of data.  Error estimates were obtained from a statistical analysis of subsets of the total data set. These represent the standard error in the mean with a $90\%$ confidence interval. Values of the reduced Soret cofficient $S^{*}_{T,1}= S_{T,1}\epsilon_{11} k_{B}^{-1} $ are given in parentheses. Values for
$\gamma_{2}^{S}-\gamma_{1}^{S}$ are also shown.
}
\begin{tabular} {|c|c|c|}
\hline
$x_{1}$ & $ S_{T,1}\times 10^{3}$ K$^{-1}$ ($S_{T}^{*}$) & $\gamma_{2}^{S}-\gamma_{1}^{S}$\\\hline 
$0.10$ & 1.499 $ \pm 0.284$ (0.180)  & 0.204\\\hline
$0.20$ &  0.953 $ \pm 0.159$ (0.114) & 0.130\\\hline
$0.30$ &  0.739 $ \pm 0.159$ (0.089)& 0.097 \\\hline
$0.40$ &  0.474 $ \pm 0.205 $ (0.057) & 0.048\\\hline
$0.50$ &  $0.0  $  & 0.0 \\\hline
\end{tabular}
\label{table3}
\end{center}
\end{table}

\begin{table}
\begin{center}
\caption{Values of $S_{T,1}$ obtained from steady-state calculations for the system
with $k_{12}=1.35$ and $\psi_{\epsilon}=1.80$ as a function of composition $x_{1}$. Results were obtained from a large number of independent simulations each with $S_{T,1}$ averaged over $330$ps of data. For each concentration  at least $1120$ independent calculations were use to obtain $S_{T,1}$. Error estimates were obtained from a statistical analysis of subsets of the total data set.  These represent the standard error in the mean with a $90\%$ confidence interval. Values of the reduced Soret cofficient $S^{*}_{T,1}= S_{T,1}\epsilon_{11} k_{B}^{-1} $ are given in parentheses. Values for
$\gamma_{2}^{S}-\gamma_{1}^{S}$ are also shown.
}
\begin{tabular} {|c|c|c|}
\hline
$x_{1}$ & $ S_{T,1}\times 10^{3}$ K$^{-1}$  ($S_{T}^{*}$) & $\gamma_{2}^{S}-\gamma_{1}^{S}$ \\\hline 
$0.10$ & $-0.180 \pm 0.105$  (-0.022) & -0.027 \\\hline
$0.20$ &  $-0.310 \pm 0.116$ (-0.037)& -0.064 \\\hline
$0.30$ &  $-0.618\pm 0.121$   (-0.074)&  -0.102  \\\hline
$0.40$ &  $-0.740\pm 0.128$  (-0.089) & -0.143 \\\hline
$0.50$ &  $-1.211 \pm 0.093 $ (-0.145) & -0.189\\\hline
$0.60$ &  $-1.406 \pm 0.149 $(-0.169) & -0.236 \\\hline
$0.70$ &  $-1.764\pm 0.118$  (-0.212)& -0.300 \\\hline
$0.80$ &  $-2.139 \pm 0.243$ (-0.257) &  -0.375\\\hline
$0.90$ &  $-2.800 \pm 0.197$ (-0.336) & -0.461  \\\hline
\end{tabular}
\label{table4}
\end{center}
\end{table}

\begin{table}
\begin{center}
\caption{Values of $S_{T,1}$ obtained from steady-state calculations for the system
with $k_{12}=1.35$ and $\psi_{\epsilon}=1.80$ for concentration $x_{1}=0.30$. Results are obtained for different values of the perturbation parameter $b$. Error bars represent the standard error in the mean with a $90\%$ confidence interval.
}
\begin{tabular} {|c|c|}
\hline
$b$ & $S_{T,1}\times 10^{3}$ K$^{-1}$ \\\hline
$5 \times 10^{-6}$ & $-0.723 \pm 0.771$ \\\hline
$1 \times 10^{-5}$ & $-0.559 \pm 0.36$ \\\hline
$2 \times 10^{-5}$ & $-0.618 \pm 0.11$ \\\hline
$4 \times 10^{-5}$ & $-0.451 \pm 0.05$ \\\hline
\end{tabular}
\label{table5}
\end{center}
\end{table}

In summary, the picture developed in our previous study of binary equimolar LJ mixtures, namely that the Soret effect can be understood from the relative resistance of the two components to accumulation in the high-density, low-temperature region, is further validated from simulations on non-ideal liquid mixtures as a function of $x_{1}$. The concentration dependence of $S_{T,1}$ reported in Ref. \cite{Artola:2007aa}due to the so-called ``chemical contribution'' appears to have a very simple explanation. Several authors, as discussed later, have touched upon this physical picture previously. In the next section, we define another quantity which is even simpler that appears to also correlate with $S_{T,1}$. The discussion is related to a measure of the relative volatility of the two components with a parallel to Raoult's law for vapor pressures and deviations due to activity coefficients which differ from unity. 

\section{Analysis based on deviations from ideal mixtures}
While the sensitivity of partial pressures to density gradients reflected in term $L_{2}-L_{1}$ is key to understanding the physical mechanism of the Soret effect, there is another analysis approach that shares much in common with activity coefficients and deviations from Raoult's law in non-ideal mixtures. This approach
is based on our previously-obtained insight that in equimolar binary liquids the component with the largest partial pressure exhibits enhanced concentration in the high-temperature region, while the component with the lowest partial pressure exhibits enhanced concentration in the low-temperature region\cite{Schelling_2023}. This picture is overly simplified and clearly cannot be applied to concentrations away from $x_{1}=0.5$. We first begin with a brief reminder of Raoult's law and how deviations in non-ideal mixtures are observed and understood. Next we develop a related, but slightly different, concept of ideal behavior, and demonstrate the key quantity that correlates with the Soret effect and $S_{T,1}$.

Raoult's law describes the vapor pressure of liquid mixtures liquid based on the mole fraction of each component and the vapor pressure of the pure components. In ideal solutions, Raoult's law states that the partial pressure $p_{s}$ of component $s$ is given by,
\begin{equation}
p_{s} = x_{s}p^{\sigma}_{s} \text{  ,}
\end{equation}
in which $p^{\sigma}_{s}$ is the vapor pressure of the pure component and $ x_{s}$ is the mole fraction of component $s$. In fact, 
Raoult's law applies to the fugacity rather than partial pressure, but the above expression applies approximately if the vapor phase can be considered to be nearly ideal.  For a two component liquid mixture,
the total vapor pressure $p$ is predicted by Raoult's law to be,
\begin{equation} 
p = x_{1} p^{\sigma}_{1} + (1-x_{1}) p^{\sigma}_{2}
\end{equation}
Deviations from Raoult's law are described in terms of activity coefficients $\gamma_{s}$ so that the total vapor pressure is given by,
\begin{equation}\label{rl}
p = x_{1}  \gamma_{1} p^{\sigma}_{1} + (1-x_{1}) \gamma_{2} p^{\sigma}_{2}
\end{equation}
Positive deviations from Raoult's law occur when the activity coefficients $\gamma_{s}>1$. This behavior occurs when intermolecular interactions between species $s=1$ and $s=2$ are relatively weak. In this case, the liquid mixture is more volatile than the prediction based on Raoult's law. Negative deviations from Raoult's law occur when the activity coefficients $\gamma_{s}<1$. This is correlated to relatively strong interaction between species $s=1$ and $s=2$. In this case, the liquid mixture is less volatile than the prediction based on Raoult's law.

Here we develop a similar picture concept of an ideal mixture, and use deviations from ideality to predict the sign of $S_{T,1}$ and at least correlate to the magnitude of $S_{T,1}$. While the simulations in this paper correspond to supercritical fluids, the same expressions should apply equally to liquids. As described previously, for each system,  conditions were adjusted across the composition range such that the total pressure of the fluid mixtures were identical. We define an ideal mixture as one in which the partial pressures can be predicted by the total pressure $p$ and the mole fractions $x_{1}$ and $1-x_{1}$. These ideal values are given by,
\begin{equation}
p_{1}^{ideal} = x_{1} p 
\end{equation}
\begin{equation}
p_{2}^{ideal} = (1-x_{1})p 
\end{equation}
The ideal case corresponds to parameters $k_{12}=1$ and $\psi_{\epsilon}=1$ which corresponds to two identical components in the mixture. Hence, the ideal reference is when the two components $s=1$ and $s=2$ are identical, and then we expect $S_{T,1}=0$ (the exception is the isotope effect which is discussion in the last section). When $k_{12}\ne 1$ and/or $\psi_{\epsilon} \ne 1$, then deviations from the ideal behavior as described above can be described by coefficients $\gamma_{1}^{S}$ and $\gamma_{2}^{S}$ defined by,
\begin{equation} \label{gs1}
p_{1}= x_{1}\gamma_{1}^{S} p 
\end{equation}
\begin{equation} \label{gs2}
p_{2} = (1-x_{1})\gamma_{2}^{S}p 
\end{equation}
The parameters $\gamma_{s}^{S}$ are clearly related to activity coefficients, but are distinct since the definition of an ideal mixture is different from what is used in Raoult's law.
However, in the case where $\psi_{\epsilon}=1$,  $\gamma_{s}^{S}$ are indeed equivalent to activity coefficients. When $\psi_{\epsilon}\ne 1$, then $\gamma_{s}^{S}$ are not equal to the activity coefficients. In the Conclusions section, the explicit connection between $\gamma^{S}_{s}$ and $\gamma_{s}$ is established with the suggestion that measurements of partial pressures in the vapor could be directly connected to $S_{T,1}$ for the liquid phase. This experimental connection does not apply to the case of supercritical fluids like those simulated here.

Now the connection between the parameters $\gamma_{1}^{S}$ and $\gamma_{2}^{S}$ and the Soret effect can be made. First notice that there is a constraint,
\begin{equation} \label{const}
x_{1}\gamma_{1}^{S}+(1-x_{1} )\gamma_{2}^{S}=1
\end{equation}
Then when $\gamma_{1}^{S} > 1$, for example, it must be that $\gamma_{2}^{S} < 1$. The simulation results demonstrate that when component $s$ has $\gamma_{s}^{S}>1$, it  will be enriched in the high-temperature region, and then the other component $s^{\prime}$ will have $\gamma_{s^\prime}^{S}<1$ and it will be enriched in the low-temperature region. 

In some sense this picture is reminiscent of deviations from Raoult's law. The parameter $\gamma_{2}^{S}$ is connected with the ``volatility'' of component $s$, which in this case is related to the tendency to have enhanced concentration in the low-density, high-temperature region. The high-temperature region is less dense that the low-temperature region. Moreover, $\gamma_{s}^{S}>1$, which in this picture implies greater ``volatility'' of component $s$, is connected with relatively weaker interactions with the mixture. This is similar to activity coefficients $\gamma_{s}>1$ and positive deviations from Raoult's law. When $\gamma_{s}^{S}<1$, we can think of component $s$ as less volatile, and hence with relatively stronger interactions with the mixture. This is similar to activity coefficients $\gamma_{s}<1$ and negative deviations from Raoult's law. 

\begin{figure}
\begin{centering}
\includegraphics[width=0.5\textwidth]{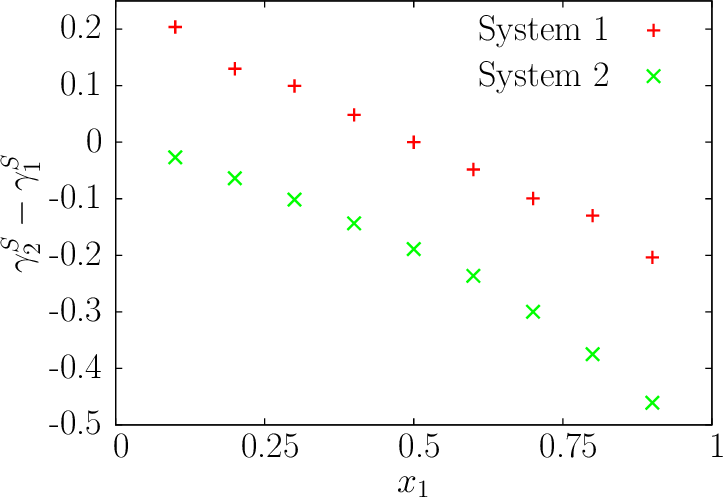} 
\caption{The quantity $\gamma_{2}^{S} - \gamma_{1}^{S}$ plotted as a function of concentration $x_{1}$. The data for ``System 1'' corresponds to  $k_{12}=1.35$ and $\psi_{\epsilon}=1.0$. The data for ``System 2'' corresponds to  $k_{12}=1.35$ and $\psi_{\epsilon}=1.80$. The results show that $\gamma_{2}^{S} - \gamma_{1}^{S}$ is not necessarily a linear function of $x_{1}$.
}
\label{fig6}
\end{centering}
\end{figure}

The computed values of $\gamma_{2}^{S}-\gamma_{1}^{S}$ are reported in Tables \ref{table3}-\ref{table4} along with the $S_{T,1}$ values.
In Fig. \ref{fig6} results for $\gamma_{2}^{S}-\gamma_{1}^{S}$ are plotted as a function of $x_{1}$. For $k_{12}=1.35$ and $\psi_{\epsilon}=1.0$, the interactions between the two components are stronger than interactions in a pure liquid, resulting in the minority component $s$ of the mixture described by $\gamma_{s}^{S}<1$. For example, when $x_{1}<0.5$, we find  $\gamma_{2}^{S}-\gamma_{1}^{S}>0$ and $S_{T,1}>0$, indicating enhanced concentration of component $s=1$ in the high-density, low-temperature region. For simulation results with $k_{12}=1.35$ and $\psi_{\epsilon}=1.80$, $\gamma_{2}^{S}-\gamma_{1}^{S} <0 $ for all concentrations simulated. This correlates with $S_{T,1}<0$ as shown in Table \ref{table4} and hence the tendency of component $s=1$ to exhibit enhanced concentration in the high-temperature region, and component $s=2$ to exhibit enhanced concentration in the low-temperature region. It is also evident that $\gamma_{2}^{S}-\gamma_{1}^{S}$ depends nonlinearly on $x_{1}$ at least for $k_{12}=1.35$ and $\psi_{\epsilon}=1.80$ results.

\begin{figure}
\begin{centering}
\includegraphics[width=0.5\textwidth]{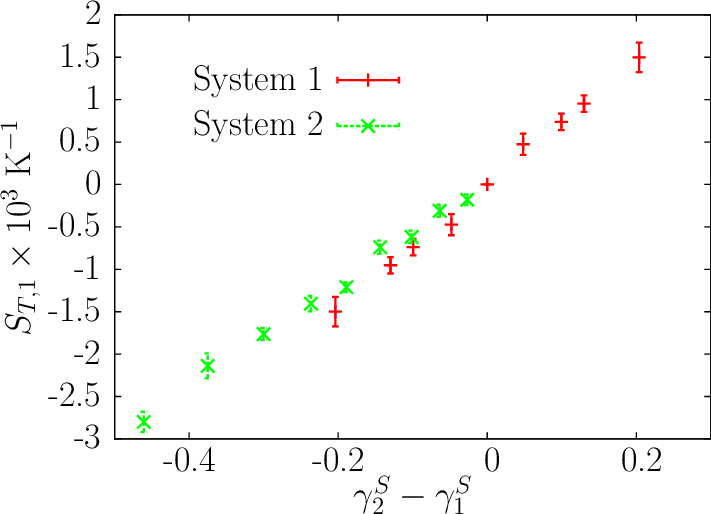} 
\caption{Soret coefficients $S_{T,1}$ plotted as a function of the difference $\gamma_{2}^{S}-\gamma_{1}^{S}$. The data points for ``System 1'' correspond
to potential parameters $k_{12}=1.35$ and $\psi_{\epsilon}=1.0$. The data points for ``System 2'' correspond
to potential parameters $k_{12}=1.35$ and $\psi_{\epsilon}=1.80$. For ``System 1'' results, due to the symmetry with respect to particles $s=1,2$, results were only computed for $x_{1}$ from $0.10$ to $0.40$. Values for $x_{1}>0.50$ were obtained from those results simply by changing the sign of $S_{T,1}$. 
}
\label{fig7}
\end{centering}
\end{figure}

In Fig. \ref{fig7}, the values of $S_{T,1}$ are plotted as a function of $\gamma_{2}^{S}-\gamma_{1}^{S}$ for both systems demonstrating the very strong correlation. For the data in this paper, $\gamma_{2}^{S}-\gamma_{1}^{S}$ determines both the sign and magnitude of $S_{T,1}$ with essentially linear dependence.
It seems straightforward why $\gamma_{s}^{S}$ should be correlated with the sensitivity of the components to density variations. When component $s$ is less volatile ($\gamma_{s}^{S}<1$), it interacts more strongly with the mixture and hence tends to exhibit enhanced concentration in regions of higher density. By contrast, when component $s$ is more volatile ($\gamma_{s}^{S}>1$), it interacts more weakly with the mixture and hence is more readily displaced from high to low-density regions. 

Finally, it is apparent that the correlation of $\gamma_{2}^{S}-\gamma_{1}^{S}$ with $S_{T,1}$ is more robust than the correlation between $L_{2}-L_{1}$ and $S_{T,1}$. This is true for small concentrations $x_{1}$ where $S_{T,1}$ has small but clearly negative values for $k_{12}$ and $\psi_{\epsilon}=1.80$.  A prediction based solely on $L_{2}-L_{1}$ for these particular cases suggests that $S_{T,1}$ will become positive at low enough concentrations, which contradicts the direct calculations of $S_{T,1}$ reported here. Nevertheless, the analysis based on pressure gradients provides a direct physical perspective for understanding the Soret effect. It is possible that
$L^{T}_{2}-L^{T}_{1}$ is plays a role that is relatively significant when $S_{T,1}$ is small. While small, these terms must play some role, and perhaps in cases where $S_{T,1}$ is small only considering the connection between $L_{2}-L_{1}$ and $S_{T,1}$ may result in an incorrect qualitative picture.

In the next section, we will explore the application of the above ideas to dense liquids substantially below the critical point. There we show that while the above quantities correlate with the sign of $S_{T,1}$, there is also a strong dependence on concentration dependence of the compressibility (or alternately the bulk modulus) of the liquid.

\begin{table}
\begin{center}
\caption{Computed values for $\gamma_{1}^{S}$,  $\gamma_{2}^{S}$, and the difference $\gamma_{2}^{S}-\gamma_{1}^{S}$ for
 the system studied by Gittus and Bresme in Ref. \cite{Gittus_2023}. The difference $L_{2}-L_{1}$ computed using the data in Table 7 are also given. These calculations correspond to reduced pressure $p^{*}=0.46$ and reduced temperature $T^{*}=0.62$. The interaction energy parameters were related by ${\epsilon_{11} \over \epsilon_{22}}=0.6$.
}
\begin{tabular} {|c|c|c|c|c|}
\hline
$x_{1}$ & $\gamma_{1}^{S}$ & $\gamma_{2}^{S}$ & $\gamma_{2}^{S}-\gamma_{1}^{S}$  & $L_{2}-L_{1}$  \\\hline
$0.10$ & 2.528 & 0.830 &  -1.698 & -2.202\\\hline
$0.20$ & 2.322 & 0.670 &  -1.652 & -1.952\\\hline
$0.30$ & 2.109 & 0.525 &  -1.585 & -1.632\\\hline
$0.40$ &1.878 & 0.415 &  -1.464 & -1.643 \\\hline
$0.50$ &1.701 & 0.299 &  -1.402 & -1.391\\\hline
$0.60$ &1.517 & 0.224 &  -1.293 & -1.360\\\hline
$0.70$ &1.370 & 0.137 &  -1.233 & -1.309\\\hline
$0.80$ &1.235 & 0.060 &  -1.175 & -1.343\\\hline
$0.90$ &1.108 & 0.023 & -1.086  & -1.132\\\hline
\end{tabular}
\label{table6}
\end{center}
\end{table}

\begin{table}
\begin{center}
\caption{Tabulated results for quantities relevant to generalized thermodynamic forces for simulations corresponding to the results
due to Gittus and Bresme\cite{Gittus_2023}. These calculations correspond to reduced pressure $p^{*}=0.46$ and reduced temperature $T^{*}=0.62$. The interaction energy parameters were related by ${\epsilon_{11} \over \epsilon_{22}}=0.6$.
}
\begin{tabular} {|c|c|c|c|c|c|c|c|c|}
\hline
$x_{1}$ & $ p_{1}$ & $p_{2} $ & $ \left({\partial p_{1} \over \partial T^{*}}\right)_{n_1,n_{2}}$ &   $ \left({\partial p_{2} \over \partial T^{*}}\right)_{n_1,n_{2}}$   & $\left({\partial p_{1} \over \partial n_{1}}\right)_{T^{*},n_{2}}$   & $\left({\partial p_{2} \over \partial n_{1}}\right)_{T^{*},n_{2}}$ &  $\left({\partial p_{1} \over \partial n_{2}}\right)_{T^{*},n_{1}}$ &   $\left({\partial p_{2} \over \partial n_{2}}\right)_{T^{*},n_{1}}$  \\\hline 
$0.10$ & 0.117 & 0.350 & 0.636 & 5.885 & 3.236 & 14.593 & 1.707 & 13.776     \\\hline
$0.20$ & 0.215 & 0.248 & 1.221 & 5.078 & 4.631 &12.309 & 2.759 & 10.636    \\\hline
$0.30$ & 0.290 & 0.168 & 1.723 & 4.140 & 5.014 & 8.733  & 3.698 &  8.267  \\\hline
$0.40$ & 0.348 & 0.115 & 2.214 & 3.301 & 5.546 & 6.686  & 4.325 &  5.933   \\\hline
$0.50$ & 0.388 & 0.068 & 2.545 & 2.646 & 5.937 & 4.603  & 4.657 &  4.600 \\\hline
$0.60$ & 0.419 & 0.041 & 2.814 & 1.872 & 6.065 & 3.142  & 4.627 &  3.077  \\\hline
$0.70$ & 0.443 & 0.019 & 3.060 & 1.346 & 6.127 & 2.112  & 4.779 &  1.939  \\\hline
$0.80$ & 0.453 & 0.005 & 3.647 & 0.922 & 6.094 & 1.206  & 4.825 &  1.132  \\\hline
$0.90$ & 0.458 & 0.001 & 2.774 & 0.304 & 5.712 & 0.523  & 4.636 &  0.387 \\\hline
\end{tabular}
\label{table7}
\end{center}
\end{table}

\section{Liquid mixtures and $S_{T,1}$ minima}
Using MD simulation of simple two-component LJ mixtures, it has recently been demonstrated that a minimum $S_{T,1}$ can be observed as a function of concentration $x_{1}$\cite{Gittus_2023}.  Since $S_{T,1}$ in Ref \cite{Gittus_2023} was negative, the magnitude of $S_{T,1}$ actually displays a maxima near $x_{1}=0.5$.  Here without validating the previous calculations of $S_{T,1}$ in Ref \cite{Gittus_2023}, we consider the results in Ref \cite{Gittus_2023} based on the same analysis used for the supercritical fluid simulations.

\begin{figure}
\begin{centering}
\includegraphics[width=0.5\textwidth]{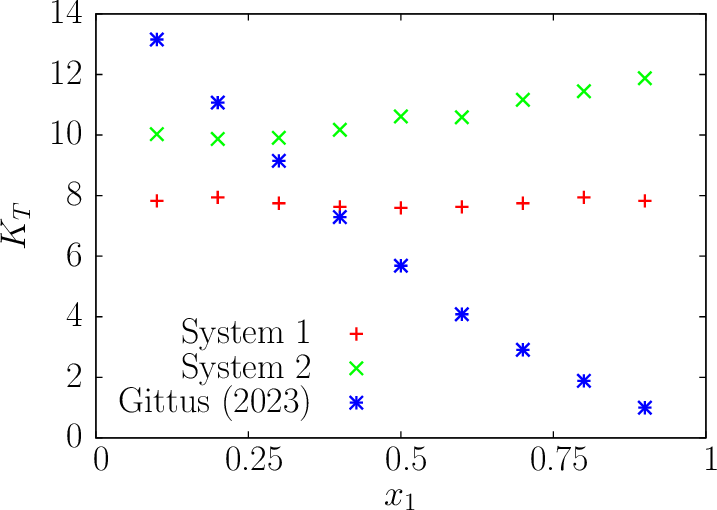} 
\caption{Computed bulk moduli $K_{T}$ in reduced units plotted as a function of concentration $x_{1}$.  The data points for ``System 1'' correspond
to potential parameters $k_{12}=1.35$ and $\psi_{\epsilon}=1.0$. The data points for ``System 2'' correspond
to potential parameters $k_{12}=1.35$ and $\psi_{\epsilon}=1.80$.  The data labeled Gittus (2023)  is for the parameters used by Gittus and Bresme in Ref. \cite{Gittus_2023}.
}
\label{fig8}
\end{centering}
\end{figure}

Following Ref. \cite{Gittus_2023}, the two components have different interaction energy parameters given by the ratio ${\epsilon_{11}\over \epsilon_{22}}=0.6$ with cross interactions defined by the Lorenz-Berthelot mixing rule. Using the notation applied previously, this corresponds to $\psi_{\epsilon} = {5 \over 3} $ and $k_{12}=1.0$. The simulations correspond to constant reduced pressure $p^{*}=0.46$ and reduced temperature $T^{*}={k_{B}T \over \epsilon_{22}}=0.62$. The mass and size parameters were the same for both components, $m_{1}=m_{2}$ and $\sigma_{11}=\sigma_{22}$.
In Table \ref{table6} values for $\gamma_{1}^{S}$, $\gamma_{2}^{S}$, and the difference $\gamma_{2}^{S}-\gamma_{1}^{S}$ are given. The values for $L_{2}-L_{1}$ are also reported in Table \ref{table6}. For each case, the quantity  $\gamma_{2}^{S}-\gamma_{1}^{S}$ most closely associated with the driving force is negative, which corresponds with the results in \cite{Gittus_2023} that $S_{T,1}<0$ for each composition. Hence the connection is demonstrated between the tendency of the most volatile component to be driven towards the high-temperature region, and the less volatile component to be driven to the low-temperature region. However, $\gamma_{2}^{S}-\gamma_{1}^{S}$ does not correlate with the magnitude of $S_{T,1}$, and moreover does not demonstrate a  connection with a minimum value for $S_{T,1}$ near $x_{1}=0.5$. This observation indicates additional factors need to be considered. It should be noted that the minimum in $S_{T,1}$ is explained by Gittus and Bresme based on a minimum in the computed thermodynamic factor. Here we provide another perspective on the minimum in $S_{T,1}$, and also present a possible reason 
no minima was found in Ref.\cite{Artola:2007aa}.

Apart from the fact that the simulations by Gittus and Bresme\cite{Gittus_2023} are for a liquid below the critical point,  the most important distinction between their work and that by Artola and Rousseau\cite{Artola:2007aa} is that the latter used the ``one-fluid'' approximation.  Within the approximation, the relative state point remains essentially fixed. To achieve that, as we explained previously, as composition is varied, the temperature $T^{*}$ is also varied. By contrast, Gittus and Bresme use the same reduced temperature $T^{*}=0.62$ as defined with respect to $\epsilon_{22}$. Consequently, $p$ and $T$ are the same for each composition in Ref.  \cite{Gittus_2023}. In that latter work, at smaller values of $x_{1}$ the system actually enters into a liquid-solid coexistence region although the system remains as a metastable liquid mixture during an MD simulation\cite{Gittus_2023}.

To further explore this point, in Table \ref{table7} we present the same quantities investigated for the Artola and Rousseau data. Namely, we present results to show how partial pressures change with temperature and densities $n_{1}$ and $n_{2}$.  The first notable result is that the partial pressures are much more sensitive to $n_{1}$ in comparison to $n_{2}$. This is the fundamental reason why $S_{T,1}<0$, resulting in enrichment in component 2 in the dense, low temperature region. Analysis of $L_{2}-L_{1}$ and $L^{T}_{2}-L^{T}_{1}$ (not shown) does not reveal any connection to a minimum value of $S_{T,1}$. Values for $L_{2}-L_{1}$, obtained from the data in Table \ref{table7}, are given in Table \ref{table6}. The negative values of $L_{2}-L_{1}$ is consistent with $S_{T,1}<0$. However, the magnitude of $L_{2}-L_{1}$  increases with decreasing $x_{1}$ without showing an extrema near $x_{1}=0.5$. 

The most remarkable difference between the results analyzed based on the Artola and Rousseau calculations\cite{Artola:2007aa} and those for liquids by Gittus and Bresme\cite{Gittus_2023} is the very different composition dependence of the isothermal bulk modulus $K_{T}$. For a fixed concentration $x_{1}$ and temperature $T$, this quantity is obtained from the data in Tables \ref{table1}-\ref{table2} and Table \ref{table7} from,
\begin{equation}
K_{T}=n \left({\partial p \over \partial n}\right)_{x_{1},T}= n x_{1}\left[\left( {\partial p_{1} \over \partial n_{1}}\right)_{n_{2},T}+\left( {\partial p_{2} \over \partial n_{1}}\right)_{n_{2},T}\right]+ n (1-x_{1})\left[\left( {\partial p_{1} \over \partial n_{2}}\right)_{n_{1},T}+\left( {\partial p_{2} \over \partial n_{2}}\right)_{n_{1},T}\right] \text{  ,}
\end{equation}
in which $n=n_{1}+n_{2}$ and $p=p_{1}+p_{2}$. Also notice that $n_{1}=n x_{1}$ and $n_{2}=n(1-x_{1})$. The results for $K_{T}$ as a function of concentration $x_{1}$ is shown for both systems from Artola and Rousseau\cite{Artola:2007aa} (Tables \ref{table1}-\ref{table2}) and the more recent results from Gittus and Bresme (Table \ref{table7}) are shown in Fig. \ref{fig8}. The striking difference is the very strong composition dependence of $K_{T}$ using the parameters from Ref. \cite{Gittus_2023} in contrast to the very weak dependence for the simulations in the one-fluid approximation as reported here and in Ref. \cite{Artola:2007aa}. In the liquid simulations of Ref. \cite{Gittus_2023}, as $x_{1}$ decreases below $0.5$, the system enters a solid-liquid coexistence region. While the system is remains a metastable liquid, this region corresponds to a more incompressible liquid, with large values of $K_{T}$ and also smaller values for the self-diffusion constants\cite{Gittus_2023}. It is the very strong dependence of $K_{T}$ on $x_{1}$ that we believe plays a role in the existence of a minimum $S_{T,1}$ near $x_{1}=0.5$. Specifically, while our results show that $\gamma_{S}^{2}-\gamma_{S}^{1}$ increases in magnitude as $x_{1}$ decreases, corresponding to a stronger driving force for separation, this is compensated by the increasing incompressibility of the liquid as $x_{1}$ decreases. 

Hence, these results suggest that the strong correlation between $\gamma_{S}^{2}-\gamma_{S}^{1}$ and $S_{T,1}$ seen in the one-fluid approximation fluids studied here and by Artola and Rousseau \cite{Artola:2007aa} is obtained in part because $K_{T}$ is relatively insensitive to $x_{1}$. This is certainly due to the fact that the one-fluid approximation attempts to maintain a consistent state point even as $x_{1}$ changes. By contrast, for the liquid simulations reported in Ref. \cite{Gittus_2023}, two competing factors result in $S_{T,1}$ taking on the largest minimum value near $x_{1}=0.5$. The first factor is that the driving force for separation becomes larger, as suggested by $\gamma_{S}^{2}-\gamma_{S}^{1}$. The second competing factor is the increase in $K_{T}$ with decreasing $x_{1}$ which tends to result in less separation.  Specifically, larger values of $K_{T}$ mean smaller density gradients result from an imposed heat source.

The above arguments can be further understood by considering $L_{2}-L_{1}$ as a function of concentration in relation to Eq. \ref{dvdt}. While an increase of the magnitude of $L_{2}-L_{1}$ should be correlated with a greater tendency towards separation, the driving force for this effect also depends on the total density gradient. This is reflected by the quantity $\tilde{n}(k,t)$ in Eq. \ref{dvdt}. The dependence of this quantity on the temperature gradient depends in part on the bulk modulus. In the system studied by Gittus and Bresme \cite{Gittus_2023}, as $x_{1}$ decreases, the system becomes more incompressible. This competing effect will tend to decrease the driving force for separation as $x_{1}$ decreases.

While these points remain to be further substantiated, they suggest additional insight into the results in Ref. \cite{Gittus_2023}, and also a potential reason why an extrema in $S_{T,1}$ was not observed by Artola and Roussea\cite{Artola:2007aa}. We also note that the microscopic picture based on a minima of the thermodynamic factor presented in Ref. \cite{Gittus_2023} is not in contradiction to the hypotheses presented here. However, it remains to be demonstrated what connection might exist between these two different perspectives.

\section{Conclusions}
It is shown here that volatility and partial pressure play a strongly determinative role in both the sign and magnitude of the $S_{T,1}$. This basic picture is identical to the previous work\cite{Schelling_2023}. Going beyond that work,
composition dependence, which can lead to changes in the sign and magnitude of $S_{T,1}$, was elucidated. Finally, a measure of volatility  $\gamma_{s}^{S}$ was defined which can be correlated to the sign and magnitude of $S_{T,1}$. This parameter is distinct from, but closely related to, the activity coefficient $\gamma_s$ of a non-ideal mixture. The physical insight obtained here is similar to how one can understand deviations from predictions of Raoult's law for the vapor pressure of a liquid mixture. We believe that this  provides useful new insight into the `chemical contribution''.

The connection between the coefficients $\gamma_{s}^{S}$ and the activity coefficients $\gamma_{s}$ can be established. Here we discuss the relevance to liquids in equilibrium with a vapor phase. The argument here applies specifically two a two-component liquid where the vapor pressures of the pure states $p_{s}^{\sigma}$ and activity coefficients $\gamma_{s}$ can be experimentally measured. We note that partial pressures can be defined either using Eq. \ref{rl} or in terms of $\gamma_{s}^{S}$ using Eqs. \ref{gs1}-\ref{gs2}. Equating these expression, we have for both components $s=1,2$,
\begin{equation}
\gamma_{s}p_{s}^{\sigma} =\gamma_{s}^{S} p
\end{equation}
Combining these two equations with the constraint Eq. \ref{const} yields expressions for $\gamma_{1}^{S}$ and $\gamma_{2}^{S}$, 
\begin{equation} \label{c1}
\gamma_{1}^{S} = {\gamma_{1} p_{1}^{\sigma} \over x_{1} \gamma_{1} p_{1}^{\sigma} +(1-x_{1}) \gamma_{2} p_{2}^{\sigma}} 
\end{equation}
\begin{equation} \label{c2}
\gamma_{2}^{S} = {\gamma_{2} p_{2}^{\sigma} \over x_{1} \gamma_{1} p_{1}^{\sigma} +(1-x_{1}) \gamma_{2} p_{2}^{\sigma}} 
\end{equation}
Hence the difference $\gamma_{2}^{S}-\gamma_{1}^{S}$, which is correlated to $S_{T,1}$, can be determined entirely in terms of quantities that can be determined from  measurable quantities of the equilibrium vapor phase.  It should be noted that this derivation neglects differences between vapor pressure and fugacity, as mentioned previously. It also neglects differences in the equilibrium compositions of liquid and vapor phases. Thus Eq. \ref{c1} is a first-order approximation. Deviations become more important for cases with strongly non-ideal vapor and for fluids that are mixtures between components with very different volatility. 

The expressions Eq. \ref{c1}-\ref{c2} indicate that the dominant factor for $S_{T,1}$ is the partial pressures $p_{s}^{\sigma}$ of the pure phases. However, the activity coefficients $\gamma_{s}$, which are themselves dependent on composition $x_{1}$, also determine $S_{T,1}$ for nonideal liquids. Specifically, in the case of an applied temperature gradient, the component with the largest value $\gamma_{s}p_{s}^{\sigma}$ is enriched in the low-density, high-temperature region, and the component with the smallest value $\gamma_{s}p_{s}^{\sigma}$ is enriched in the high-density, low-temperature region.  

We also note that liquid mixtures in equilibrium with their vapor exhibit behavior that can be related to the picture above. Specifically, for a two-component liquid, the vapor will be enriched in the component with the largest value $\gamma_{s}p_{s}^{\sigma}$, and the solution will be enriched in the component with the smallest value $\gamma_{s}p_{s}^{\sigma}$. This well-known result is the physical basis for distillation. We suggest that this is related to the Soret effect, even though the Soret effect is a non-equilibrium phenomenon. 

Analysis of the recent work of Gittus and Bresme \cite{Gittus_2023} for liquid mixtures indicates that the additional factors need to be considered when trends are not determined using the one-fluid approximation. It was shown that while $\gamma_{2}^{S}-\gamma_{1}^{1}$ and $L_{2}-L_{1}$ determine the sign of $S_{T,1}$, the correlation between these quantities and the magnitude of $S_{T,1}$ depends on other factors. Specifically, it is suggested that the driving force for separation depends on the total density gradient which depends in part on the bulk modulus. In Ref. \cite{Gittus_2023}, we show that the liquid becomes increasingly incompressible as concentration $x_{1}$ decreases. We argue that this effect is partially responsible for the local minimum value of $S_{T,1}$ near $x_{1}=0.5$\cite{Gittus_2023}. In short, the dependence of the compressibility of a liquid on concentration also is predicted to play a role in the magnitude of $S_{T,1}$.

It should be possible for a broad review of previous results based on the insight developed here to determine whether this picture broadly holds. In some of our own previous results, this was indeed possible. Specifically, in our paper on liquid metal alloys\cite{TUCKER201654}, partial pressures for both components were reported along with Green-Kubo calculations of the heat of diffusion which is closely connected to the Soret coefficient. Results for two different potentials for equimolar NiAl liquids were reported which showed different signs for partial pressures and the heat of transport. Specifically, it was found that using an embedded atom model (EAM) potential that predicted $p_{Ni} > 0$ and $p_{Al}<0$, the computed heat of transport demonstrated that Ni is enriched in the high-temperature region and Al is enriched in the low-temperature region. Results using a different EAM potential found $p_{Ni} <0$ and $p_{Al}>0$, and correspondingly the opposite behavior for the segregation. These results agree qualitatively with the insight in this paper. Another EAM potential for NiCu was also reported in Ref. \cite{TUCKER201654} . In that case, both parameters had small negative partial pressures for the reported conditions, but with $p_{Cu}<p_{Ni}$, and Green-Kubo results indicated Ni enrichment in the high-temperature region, and Cu enrichment in the low-temperature region. This is again consistent with the the insight obtained in this paper, and hence  further validates, or at least does not contradict, this picture. Composition dependence was not considered in this earlier work.

The quantity $\gamma_{s}^{S}$  seems to be most closely connected to the concept of ``thermophobicity'' developed by previous researchers in understanding the Soret effect across a wide range of different liquid mixtures\cite{Hartmann:2012va,Hartmann_2014}. Here we have made a connection between $S_{T,1}$ and $\gamma_{2}^{S}-\gamma_{1}^{S}$. One novel aspect of thermophobicity is that it appears to be a concept connected to the pure fluid, which then is predictive of the behavior of mixtures.  Based on the insight here, we suggest that the most likely connection is between thermophobicity and volatility, and hence the equilibrium vapor pressure of a liquid. However, deviations would be expected depending on how the two components in a mixture interacted. These deviations should be connected to deviations from Raoult's law. It might even be possible to predict the Soret coefficients based on measurements of the vapor pressures of the pure components and measured deviations from Raoult's law and represent what has been called the ``chemical contribution''. The latter might play a role of corrections to predictions made based on thermophobicity of the pure components which would be expected to lead to concentration dependence of the Soret effect.

Some of the ideas presented here are very closely connected to those in the original work of Artola and Rousseau \cite{Artola:2007aa}, where $S_{T,1}$ was correlated to differences in partial enthalpy $h_{2}-h_{1}$ which includes the difference $p_{2}-p_{1}$. However, this previous work does not develop as clear a physical picture as the developed in this and our previous effort Ref. \cite{Schelling_2023}. Namely, that the segregation effect is generated by gradients in partial pressures which exist even when the total pressure does not exhibit a gradient. Hence, we believe that our work extends the previous observations from Ref. \cite{Artola:2007aa} which points in the right direction without developing an entirely clear microscopic picture.

There have been other mechanistic descriptions in the literature which have significant connections to this work. In the work by Morozov \cite{Morozov:2009wa}, the connection between partial pressure and the Soret effect was examined. Much of the work related to the Morozov model has been reviewed in Ref. \cite{Hoang_2022}. The overall qualitative description presented in Ref. \cite{Morozov:2009wa} is in exact agreement with our explanation of the Soret effect. Specifically, Morozov notes that the fluid is more dense in the low-temperature region, and that the concentration gradient emerges due to competition between the two components of the fluid.  The competition between particles with different interaction strengths and sizes has been noted by others\cite{Reith_2000,Artola:2007aa,Hoang_2022}. However, Morozov considers only gradients of partial pressure with respect to temperature, and not directly gradients due to fluid density. Instead, Morozov attempts to capture the effect of fluid density using differences in the ``compressibility factor'' of the solute in the solvent and the pure solvent.  While some aspects have been tested for hydrocarbon mixtures with good success\cite{Morozov:2009wa,Debuschewitz:2001aa}, there has not been systematic testing of the model. Finally, the Morozov model is applicable only to binary mixtures at infinite dilution.

In the review article by Hoang and Galliero\cite{Hoang_2022}, the general observations, accumulated primarily by MD simulation over the years, that ``small'' particles with relatively strong interactions are enriched in the cold, dense region, while ``big'' particles with relatively weak interactions are enriched in the hot, less dense region, was discussed. It was noted that no existing theories\cite{Kempers_1989,Shukla_1998,Kempers_2001,Farago_2006} are able to completely describe all of the relevant effects\cite{Hoang_2022}, although it has been noted that the model due Shukla and Firoozabadi\cite{Shukla_1998} has had the most notable successes\cite{Artola:2007aa,Hoang_2022}. The main contribution of our work is to note that, at least for spherical particles, the relevant effects can be captured by simply considering partial pressures and the relative volatility of the two components. The relative partial pressures are determined by both particle sizes and interaction strengths. Moreover, as shown above, it may be possible in the case of liquids below the critical point to measure partial pressures of the vapor as an indirect way of assessing the direction and magnitude of the Soret effect.

If the picture presented here represents the right physics to understand the Soret effect, then it should be possible to generalize to other materials systems beyond liquid mixtures, including thermodiffusion of interstitials in a solid or thermodiffusion in solid solutions, for example via a vacancy mechanism. There may be additional complications in systems with long-range Coulomb interactions, so these are not considered here further. 

First we address the case of interstitials in a lattice. In this case, the host lattice is not expected to diffuse, but rather display lattice strain associated with thermal expansion. Interstitials in a crystalline lattice can introduce either compressive or tensile stress. Large interstitial atoms are generally correlated with compressive stress. It should be expected that an interstitial component $s$ which introduces compressive stress should have $\gamma_{s}^{S}>1$ and hence should accumulate in the low-density, high-temperature region. It is possible that smaller interstitials, for example hydrogen in a metal lattice, would be associated with local tensile stress, and hence would be expected to have $\gamma_{s}^{S}<1$ indicating a tendency to accumulate in high-density, low-temperature regions. We note that we have published on the Pd:H system which showed $H$ accumulation in the high-temperature region\cite{Tucker_2013,Schelling2012ComputationalMF}. However, in that model, the Pd-H interactions were purely repulsive and H interstitials introduce a local compressive lattice strain. Thus, the picture obtained in Ref. \cite{Tucker_2013,Schelling2012ComputationalMF} is consistent with the physical mechanism proposed here, but we are not sure at all if the results in those previous papers are actually predictive of a real Pd:H system, but we believe they probably are not. For example, in addition to the purely repulsive nature of the Pd-H interaction, we have recently found that the models used in these previous papers actually predict regions of negative thermal expansion coefficients for the Pd lattice.

Next we address the case of a solid solution where interdiffusion is mediated by a vacancy mechanism. These systems are difficult to directly explore using MD simulation, since atom hops are ``rare events''. In several past works we presented a constrained-dynamics simulation approach to the thermodiffusion problem\cite{McDargh_2011,Tucker_2014,Schelling_2014}. In those calculations, the heat of transport, again closely related to the Soret coefficient, was determined for vacancies, but we did not continue with the work long enough at that time to address solid solutions. However, it might also be possible in cases where experimental thermodiffusion results exist to compute the $\gamma_{s}^{S}$ parameters for the components of the solution and identify whether or not the proposed correlations are indeed observed.   Moreover, it might be possible to compute $\gamma_{s}^{S}$ from first-principles (e.g. based on density-functional theory) and determine whether or not correlations exist with experimental data.

We briefly discuss how the so-called ``isotope effect'' fits into this picture, since this has been shown to be distinct from ``chemical contributions'' to the Soret effect. In our past work, it was shown that heat pulses result in damped compression waves. As material flows back from the dense cold region to the hot region, the lighter atoms tend to respond more quickly and accumulate to higher concentrations in the high-temperature region\cite{Schelling_2023}. In this paper, the possible role for mass differences is suggested by Eq. \ref{v12}. Therefore, development of a complete theory of the Soret effect would require both a connection to the quantities $\gamma_{s}^{S}$ and the different responses to pressure gradients depending on relative masses. It was also suggested in our past work that some of this insight might come simply from a better understanding of ``fast sound'' in liquid mixtures. A complete theory would need to include both chemical contributions based on differences in $\gamma_{s}^{S}$ for the components in a liquid mixture, as well as differences in molar mass. In addition, Eq. \ref{fs} for the driving force also suggests that the thermal expansion of the mixture is relevant, since this connects the magnitude of the temperature gradient to the density gradient. 

Lastly, others have noted an apparent correlation to the moment of inertia\cite{Debuschewitz:2001aa,Artola:2007aa}, which this paper, dealing entirely with spherical point particles, does not address. Along with the isotope effect, there are all areas subject to further investigation to develop a more comprehensive theory. 

\section{Supplementary Material}

The supplementary material contains details on the simulation conditions including reduced temperatures and densities. Tabulated results for 
data shown in Figs. 1-4 and Fig. 8 are also included.

\section{Acknowledgements}
The authors would like to acknowledge support from the UCF STOKES computational cluster. We also benefitted from discussions with Prof.
Daniel Halltinan at Florida State University, who has suggested a connection between the Soret effect and chemical activity. 

\section{Data Availability Statement}
The data that supports the findings of this article are available within the article and the supplementary material.

\newpage


\end{document}